\documentclass[%
 reprint,
superscriptaddress,
 amsmath,amssymb,
 aps,
]{revtex4-2}

\usepackage{graphicx}% Include figure files
\usepackage{dcolumn}% Align table columns on decimal point
\usepackage{bm}% bold math
\begin{document}

\preprint{APS/123-QED}

\title{Coupling trapped ions to a nanomechanical oscillator}

\author{Moritz Weegen}
\affiliation{Department of Chemistry, University of Basel, Klingelbergstrasse 80, 4056 Basel, Switzerland}%Lines break automatically or can be forced with \\
\affiliation{Swiss Nanoscience Institute, University of Basel, Klingelbergstrasse 82, 4056 Basel, Switzerland}
% \email{moritz.weegen@unibas.ch}
\author{Martino Poggio}%
\affiliation{Department of Physics, University of Basel, Klingelbergstrasse 82, 4056 Basel}
\affiliation{Swiss Nanoscience Institute, University of Basel, Klingelbergstrasse 82, 4056 Basel, Switzerland} %\email{martino.poggio@unibas.ch}
\author{Stefan Willitsch}%
\email[Corresponding author: ]{stefan.willitsch@unibas.ch}
\affiliation{Department of Chemistry, University of Basel, Klingelbergstrasse 80, 4056 Basel, Switzerland}
\affiliation{Swiss Nanoscience Institute, University of Basel, Klingelbergstrasse 82, 4056 Basel, Switzerland}

\date{\today}

\begin{abstract}
Cold ions in traps are well-established, highly controllable quantum systems with a wide variety of applications in quantum information, precision spectroscopy, clocks and chemistry. Nanomechanical oscillators are used in advanced sensing applications and for exploring the border between classical and quantum physics. Here, we report on the implementation of a hybrid system combining a metallic nanowire with laser-cooled ions in a miniaturised ion trap. We demonstrate resonant and off-resonant coupling of the two systems and the coherent motional excitation of the ion by the mechanical drive of the nanowire. The present results open up avenues for mechanically manipulating the quantum motion of trapped ions, for the development of ion-mechanical hybrid quantum systems and for the sympathetic cooling of mechanical systems by trapped ions and vice versa.
\end{abstract}


\maketitle

%\tableofcontents

Laser-cooled ions stored in harmonic traps \cite{major05a} count among the best controlled quantum systems \cite{leibfried03a}. Single ions or strings of ions can be cooled to the quantum regime of motion in the trap, thus realising quantum harmonic oscillators which find a variety of applications across different branches of sciences and technology. The ability to prepare tailored quantum states of motion and entangle them with the ions' internal degrees of freedom, as well as their excellent coherence properties \cite{wang21a}, render them a leading platform for quantum computing \cite{haeffner08a, fluehmann19a, pino21a} and quantum simulation \cite{lanyon11a, blatt12a, whitlow23a}. Similarly, trapped ions form the basis of some of today's most accurate clocks \cite{ludlow15a} and have enabled new types of ion-neutral collision and chemical experiments at very low energies \cite{willitsch12a, meir16a, willitsch17a}. Their unique properties have also been harnessed in hybrid experiments with ultracold atoms \cite{willitsch15a, tomza19a}.

\begin{figure*}%[!h]
    \centering
    \includegraphics[width=160mm]{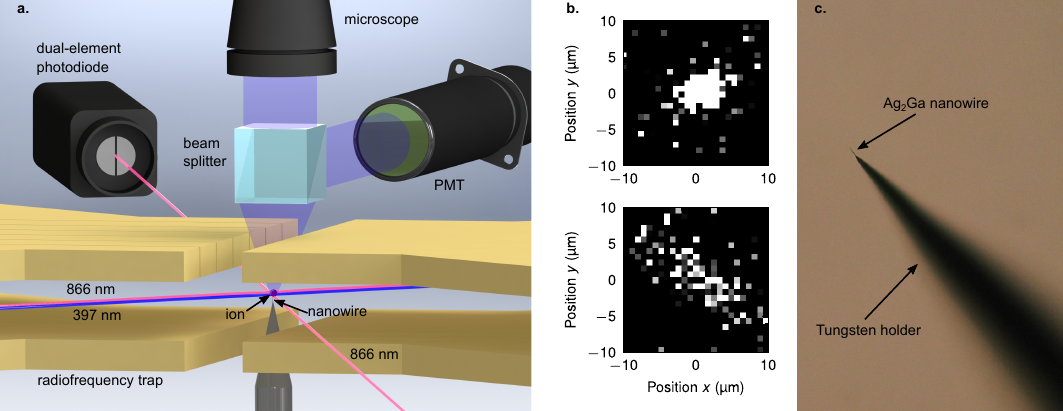}
    \caption{\textbf{Ion-nanowire hybrid system.} \textbf{a}, Schematic of the experimental setup. An Ag$_2$Ga  nanowire was positioned close to trapped $^{40}$Ca$^+$ ions in a miniaturised RF trap. The ions were cooled with laser beams of wavelengths 397~nm and 866~nm. Oscillations of the nanowire were imaged by projecting its shadow onto a dual-element photodiode with a laser beam at 866~nm. The resonance fluorescence of the ions generated during laser cooling was imaged by an EMCCD camera coupled to a microscope above the experimental chamber. A beam splitter directed half of the collected ion fluorescence to a photomultiplier tube (PMT) for the detection of driven ion motion using photon correlation. See text for further details.
    \textbf{b}, Fluorescence image of a single trapped ion at rest (top) and driven by the nanowire (bottom). \textbf{c}, Optical microscope image of the nanowire and its support at $20\times$ magnification.}
    \label{fig1}
\end{figure*}

Traditionally, trapped ions have been manipulated using laser \cite{leibfried03a}, electric \cite{meekhof96a, srinivas23a}, magnetic and microwave \cite{johanning09a,ospelkaus11a} fields. An intriguing prospect is their manipulation by other types of oscillators, e.g., other trapped ions \cite{brown11a, harlander11a, an22a} or nanomechanical systems. As explored theoretically in Refs. \cite{tian04a, hensinger05a, liu07a, nicacio13a, kotler17a, fountas19a}, ion-mechanical hybrid systems offer possibilities for extending the available techniques for the cooling, manipulation, control and readout of both constituents up to realising novel hybrid quantum systems \cite{daniilidis13a, kurizki15a} in which entanglement between ions and mechanical objects can be realised. 

In this context, the research on mechanical oscillators on the nanometer scale has progressed rapidly in recent years. Their properties as objects on the border between classical and quantum physics make them excellent candidates for the realisation of ion-mechanical hybrid systems \cite{bachtold22a}. Cryogenic cooling of nanomechanical oscillators in combination with other cooling techniques has enabled the leap from the classical to the quantum regime by preparing them close to their motional ground state \cite{oconnell10a,chan11a,teufel11a,rossi18a,tebbenjohanns21a}. They have also been successfully coupled to ultracold atoms \cite{vochezer18a,joeckel15a,hwang20a,schmid22a,verhagen12a}. 

Here, we present a hybrid system consisting of trapped, laser-cooled $^{40}$Ca$^{+}$ ions in a miniaturised linear radiofrequency (RF) ion trap coupled to a conductive nanowire \cite{biedermann10a} as previously explored theoretically in Ref. \cite{fountas19a}. 
The present hybrid system was realised with single trapped ions, strings of several ions or larger ion ensembles forming three-dimensional Coulomb crystals \cite{willitsch12a}. A bias voltage applied to the nanowire coupled the two systems by their mutual electrostatic interaction. 
Mechanically driving the oscillation of the nanowire modulated the electric field experienced by the ions leading to an effective transfer of energy when the drive was resonant with the motion of the ions in the trap.

%%%%%%%%%%%%%%%%%%%%%%%%%%%%%%%%%%%%%%%%%%%%%%%%%%%%%%%%%
% EXPERIMENT
%%%%%%%%%%%%%%%%%%%%%%%%%%%%%%%%%%%%%%%%%%%%%%%%%%%%%%%%%

Fig. \ref{fig1} shows a schematic of the experimental setup with the most important components (see Methods for details). Ions were confined and laser-cooled in a microstructured linear-quadrupole radiofrequency trap \cite{willitsch12a} consisting of electrodes in a stacked-wafer configuration. The ions were observed by imaging their resonance fluorescence generated during laser cooling onto an EMCCD camera coupled to a microscope (Fig.~\ref{fig1}b). Motional excitation of the ions was quantified using the photon correlation method \cite{berkeland98a, bwang10a, keller15a}. 

The nanomechanical oscillator used was a Ag$_2$Ga cantilever positioned at the tip of a tungsten support with $250$~\textmu m base diameter and $\approx4.8$~mm length. The length of the nanowire was measured with a microscope to be $\approx16$~\textmu m with a diameter of $\approx150$~nm as specified by the manufacturer. Fig. \ref{fig1}c shows a microscope image of the nanowire on the tungsten tip. The oscillator was mounted on three nanopositioners allowing its free deployment in all spatial directions inside the trap. The wire was mechanically driven in the direction of the longitudinal trap axis ($z$) by a piezoelectric actuator attached to the assembly. %The nanowire was positioned in close proximity ($\approx 300$~\textmu m) to the ion-trap center in order to achieve a strong coupling between the two systems.

%%%%%%%%%%%%%%%%%%%%%%%%%%%%%%%%%%%%%%%%%%%%%%%%%%%%%%%%%
% RESULTS
%%%%%%%%%%%%%%%%%%%%%%%%%%%%%%%%%%%%%%%%%%%%%%%%%%%%%%%%%

\begin{figure}%[!h]
    \centering
    \includegraphics[width=\columnwidth]{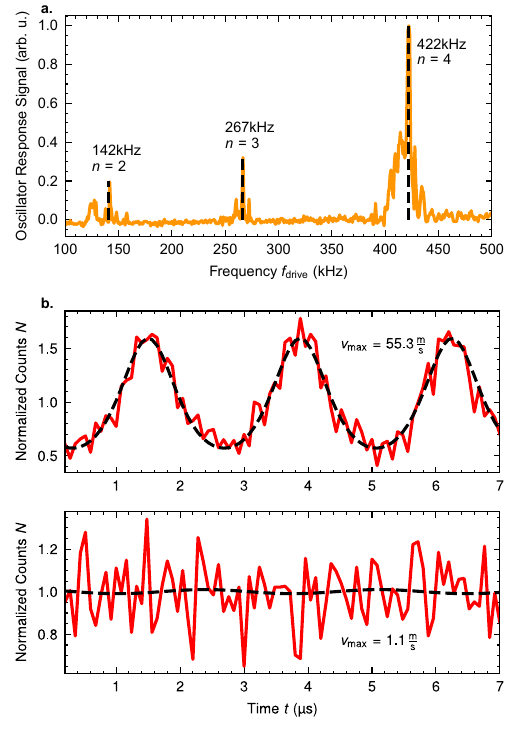}
    \caption{\textbf{Oscillations of the nanowire and the ions.} \textbf{a}, Mechanical excitation spectrum of the nanowire obtained by imaging its vibration with a dual-element photodiode. The features represent excitations of the eigenmodes $n$ of the nanowire on its support. The strong $n=4$ mode at 422~kHz was used for the excitation of the ion. Black dashed lines correspond to eigenfrequencies obtained from theoretical modelling.
    \textbf{b}, Histograms of arrival times of resonance-fluorescence photons of the ions correlated with the periodic drive signal of the nanowire under near-resonant (top) and far off-resonant (bottom) motional excitation by the mechanical oscillator. The black dashed lines show fits of the data from which the velocity amplitude $v_\text{max}$ of the ions during their oscillation was obtained (Methods).}
    \label{fig2}
\end{figure}

The eigenmodes and, therefore, also the vibration frequencies of mechanical oscillators are determined by their mass, material and geometry. To characterise the vibrational modes of the present nanowire, the drive frequency $f_{\text{drive}}$ applied to the piezo was scanned. Its oscillations were measured by imaging the tip of the oscillator placed in the waist of a focused laser beam onto a dual-element Si PIN photodiode \cite{gloppe14a, lepinay17a} (Methods). Fig. \ref{fig2}a shows a mechanical excitation spectrum thus obtained in the frequency interval from 100 to 500~kHz.

In this range, the strongest mechanical response was found around $f_{\text{drive}}=422$~kHz. Structural mechanics simulations performed with the COMSOL Multiphysics program \cite{COMSOL} indicate that this resonance corresponds to the $n=4$ eigenmode \cite{hauer13a} which represents a combined oscillation of the nanowire and its support (Methods). 

To demonstrate the coupling of the ion-nanowire system, the motional excitation of the ions by a resonant drive of the mechanical oscillator was explored. For this purpose, the piezo frequency was set to the mechanical resonance of the nanowire at 422~kHz. The oscillation frequency of the ion along the longitudinal trap axis was matched by suitably adjusting the static potentials applied to the electrodes (Methods). Close to resonance between the nanowire and ion oscillations, both systems coupled efficiently resulting in a transfer of energy from the mechanical oscillator to the ion (bottom image in Fig.~\ref{fig1}b) \cite{fountas19a}.

During driven oscillations in the trap, the ions experienced periodically varying Doppler shifts with respect to the cooling lasers modulating their resonance fluorescence. The arrival times of fluorescence photons on a photomultiplier tube (PMT) were correlated with the periodic piezo drive at $f_{\text{drive}}$ using lock-in detection. Ion velocities were then extracted from the time-resolved fluorescence profiles (Methods). Fig. \ref{fig2}b shows typical time-resolved fluorescence traces for a near-resonant ($f_{\text{drive}}=421.0$~kHz, top panel) and and far off-resonant ($f_{\text{drive}}=366.2$~kHz, bottom panel) drive of the ion motion. In these measurements, a voltage $V_{\text{nw}}=1.2$~V was applied to the nanowire which was sinusoidally driven with a voltage amplitude $V_{\text{piezo}}=5$~V applied to the piezo. The equilibrium distance between ion and nanowire was determined from a fit to experimental data (see below) as  $d\approx 300$~\textmu m.  A clear modulation of the fluorescence yield can be seen at strong excitation while the signal appears largely unmodulated with an off-resonant drive. The maximum ion velocities $v_{\text{max}}$ during oscillation were determined to be $v_\text{max}~=~55.3\pm2.0$~m/s and $1.1\pm2.2$~m/s for the examples shown at the top and bottom in Fig.~\ref{fig2}b, respectively. 

\begin{figure}
    \centering
    \includegraphics[width=\columnwidth]{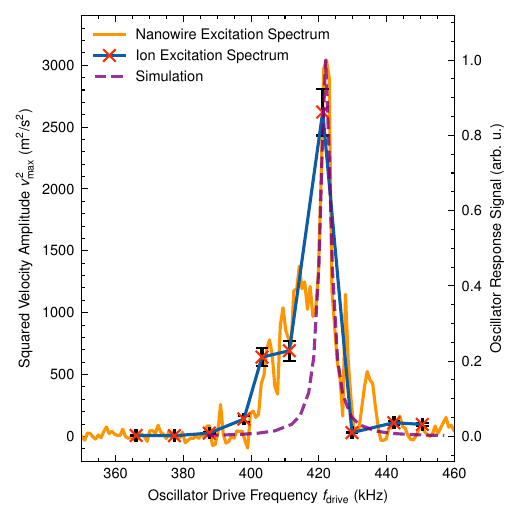}
    \caption{\textbf{Mechanical excitation of the ion by the nanowire.} Squared maximum velocities $v^2_{\text{max}}$ (red crosses) of two trapped ions over a range of drive frequencies $f_{\text{drive}}$ applied to the nanowire. The orange trace corresponds to the mechanical excitation spectrum of the nanowire shown in Fig. \ref{fig2}a. Motional excitation of the ion could only be observed around the mechanical resonance of the nanowire demonstrating the resonant coupling of the two systems. Classical dynamics simulations (purple dashed line) reproduce the frequency response of the ion excitation at the specific experimental parameters. All traces were normalised to the experimental data for comparison.  Error bars represent a combination of the errors of the fits for determining $v_\text{max}$ from time-resolved ion-fluorescence profiles and the statistical standard errors of 3 measurements.}
    \label{fig3}
\end{figure}

To confirm that the motion of the ion was indeed a consequence of the mechanical action of the nanowire rather than a spurious excitation from oscillating stray electric fields in the experiment, the frequency of the voltage applied to the piezo to drive the nanowire was scanned across the strong mechanical resonance at 422~kHz observed in Fig.~\ref{fig2}a. In each scan step, the oscillation frequency of the ion along the $z$ axis was matched to the piezo drive frequency with an accuracy of $\pm0.2$~kHz by adjusting the static trapping potentials. The motional excitation of the trapped ions was quantified as above by measuring their maximum velocities $v_{\text{max}}$ using the photon correlation method. Fig.~\ref{fig3} shows the squared maximum velocities of the ions $v_\text{max}^2$ (red crosses), which are proportional to their maximum kinetic energy in the trap, in relation to the piezo drive frequency and superimposed onto the mechanical drive spectrum of the nanowire (orange trace). The measurement was performed with a string of two ions. 

Fig.~\ref{fig3} shows that the motion of the ions is only excited when the piezo drives the motion of the mechanical oscillator. This direct correlation of the ion motion with the vibration of the nanowire proves the mechanical nature of the ion excitation and thus the successful coupling between the two systems. The dashed purple curve in Fig.~\ref{fig3} represents the results of a classical simulation of the motional excitation of a single ion by the action of the nanowire. The simulation treats the excitation of the ion by the nanooscillator by approximating the system as two oscillating point charges corresponding to the leading term in a multipole-expansion of their mutual electrostatic interactions  (Ref.~\cite{fountas19a} and Methods). While this simplified treatment cannot capture the asymmetric lineshape of the nanowire's mechanical resonance (which we attribute to overlapping mechanical modes of the entire assembly), it reproduces the general features of the frequency response of the excitation of the ion by the nanooscillator. The modelling also shows that the degree of motional excitation of the ion results from the balance of its continuous cooling by the lasers with the transfer of energy from the nanowire. 

\begin{figure}
    \centering
    \includegraphics[width=\columnwidth]{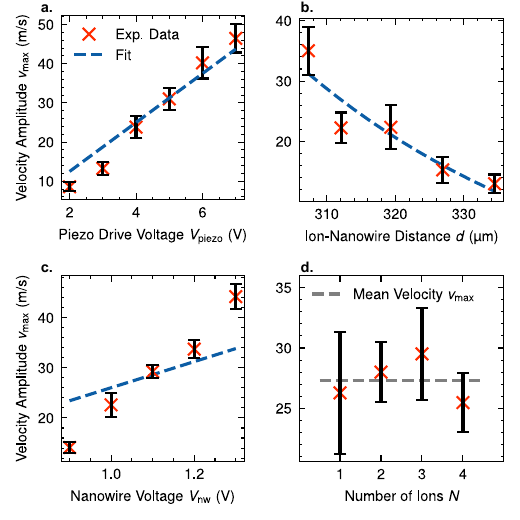}
    \caption{\textbf{Variation of coupling parameters.} Maximum ion velocities $v_\text{max}$ as a function of \textbf{a.} the voltage $V_\text{piezo}$ of the piezo actuator driving the nanowire, \textbf{b.} the ion-nanowire distance $d$, \textbf{c.} the voltage $V_\text{nw}$ applied to the nanowire, and \textbf{d.} the number of ions in the trap. Dashed lines show the results of a classical simulation of the system, see text for details. Error bars represent a combination of the errors of the fits for determining $v_\text{max}$ from time-resolved ion-fluorescence profiles and the statistical standard errors of 3 measurements.}
    \label{fig4}
\end{figure}

To further explore the properties of the coupled system, its response to variations of salient experimental parameters was studied. Fig.~\ref{fig4}a and b show the maximum velocities of the ions excited by the nanooscillator on the 422~kHz resonance as a function of the piezo drive voltage and the ion-nanowire distance, respectively. As expected, the excitation of the ions is increased with increasing the oscillation amplitude of the nanowire, i.e., with increasing piezo voltage, and decreased with increasing distance between the two subsystems. The dashed blue lines show least-squares fits of an approximated classical simulation model to the combined data. In the fit, the amplitude $A$ of the nanowire oscillation and the ion-nanowire distance $d$ were adjusted as free parameters (Methods). The good agreement illustrates that the dynamics of the system and its dependence on these specific parameters is well captured by the theory. 

Fig.~\ref{fig4}c shows the variation of the ion's maximum velocity subject to the resonant drive at varying bias voltages applied to the nanowire. The coupling, and hence the motional excitation of the ion, seem to increase linearly with the charge on the nanowire in the voltage range studied. The dashed blue line shows the results of simulations assuming the same parameters as used in Figs.~\ref{fig4}a,b. While the simulation captures the trend correctly, it predicts a weaker dependence of the ion velocity on the nanowire voltage than observed experimentally. We attribute this discrepancy to the simplifications of the model, in particular to the neglect of asymmetries in the charge distribution of the nanowire on its holder, i.e., to higher-order terms in the multipole expansion of the electrostatic interaction between the two systems, and to the possible presence of stray electric fields in the setup. 

Finally, Fig.~\ref{fig4}d shows the dependence of the maximum velocity on the number of trapped ions subjected to the drive by the nanooscillator. It can be observed that the kinetic energy imparted to each ion is independent of the other ions in the trap, consistent with the notion that the center-of-mass-motion of the entire ion crystal is excited by the nanowire. This is also supported by the ion images which show a uniform blurring of the fluorescence and hence uniform excitation of multi-ion crystals in this scenario. 

%%%%%%%%%%%%%%%%%%%%%%%%%%%%%%%%%%%%%%%%%%%%%%%%%%%%%%%%%
% DISCUSSION
%%%%%%%%%%%%%%%%%%%%%%%%%%%%%%%%%%%%%%%%%%%%%%%%%%%%%%%%%

The present study demonstrated the coupling of trapped ions to a nanomechanical oscillator in a novel hybrid system consisting of a miniaturised linear radiofrequency ion trap and a nanowire assembly. In the present experiments, the nanowire effectively acted as a miniaturised, movable, oscillating trap electrode which resonantly couples to the ions in a highly controllable fashion. Resonant excitation of the motion of trapped ions, ranging from single ions to small Coulomb crystals, was achieved in the classical regime and control over the ion motion was demonstrated by the variation of different coupling parameters.

The present results lay the basis for further experiments. As explored theoretically in Ref. \cite{fountas19a}, cooling the ion into the quantum regime of motion in the trap by, e.g., resolved-sideband methods \cite{leibfried03a} paves the way for generating different quantum states of ion motion through the action of the mechanical oscillator. These include, among others, large coherent states and multi-component Schr\"odinger-cat states which are challenging to engineer by conventional optical means \cite{franke23a}. Additionally, using nanooscillators in the quantum regime, e.g., low-mass oscillators like carbon nanotubes in a dilution-refrigerator environment, offers perspectives for realising a completely quantum hybrid system in which the exchange of discrete phonons, quantum entanglement and mutual sympathetic cooling can be realised \cite{fountas19a}. 

In this context, the key advantages of using nanomechanical oscillators are the capability for laserless state manipulation of the ions, the easy control over the mechanical drive strength as well as the possibility to quickly adjust the coupling strength on-the-fly by the change of parameters such as the nanowire's position and applied voltages. Moreover, their small size renders them attractive for integration in scalable miniaturised devices as required for large ion quantum networks \cite{pino21a}.
In addition to these prospects, the present experiment can be further developed in a variety of directions. Driving the nanowire optically may offer more localised control over its oscillation without driving other parts of the assembly such as its holder \cite{gloppe14a}. Integrated optics could improve the optical in-situ readout of the nanowire resonances. Lastly, the use of other types of nanomechanical oscillators such as membranes as candidates for the mechanical quantum-state manipulation of ions could be investigated.\\

\begin{acknowledgments}
We thank Dr. Anatoly Johnson, Georg Holderied, Philipp Kn\"opfel and Grischa Martin for technical assistance. We thank Dr. Panagiotis Fountas for his contributions to the development of the experimental setup and Dr. Adrien Poindron for useful discussions. This work is supported by the Swiss Nanoscience Institute (grant nr. P1808), the Swiss National Science Foundation (grant nr. 200021\_204123), and the University of Basel.
\end{acknowledgments}

\appendix

\section*{Methods}

\subsection*{Ion trapping and cooling}

The linear-quadrupole RF ion trap consisted of four wafers at a separation of 400~\textmu m fabricated from laser-cut aluminium oxide sputtered with gold following the layout proposed in Ref. \cite{fountas19a}. Two diagonally opposed wafers formed the RF electrodes. The remaining two wafers were sectioned into seven individually addressable segments of 400~\textmu m length to which static voltages were applied for confining the ions along the longitudinal axis of the trap and manipulating the position of the ions. Typical static voltages applied were on the order of 0.5~V to 1~V for stable ion confinement along the longitudinal trap axis at frequencies $f_z\approx 400$~kHz. The secular oscillation frequencies of the ions were measured using resonant motional excitation by additional RF fields applied to one of the trap electrodes \cite{drewsen04a}. The radiofrequency used was $f_{\text{RF}}=21.629$~MHz at an amplitude $V_{\text{RF}}\approx50$~V.\\
$^{40}$Ca$^+$ ions were loaded by photoionisation of Ca atoms emanating from an oven source using two diode laser beams at 423~nm and 375~nm inside the trap. The ions were Doppler laser-cooled on the $(4\text{s})~^2\text{S}_{1/2} \leftrightarrow (4\text{p})~^2\text{P}_{1/2} \leftrightarrow (3\text{d})~^2\text{D}_{3/2}$ system of optical cycling transitions using diode laser beams at 397~nm and 866~nm \cite{willitsch12a}. The laser beams were inserted at an angle of $\approx45^\circ$ to all principal trap axes (see Fig.~\ref{fig1}) to ensure cooling in all spatial directions. 

\subsection*{Vibrational spectrum of the nanowire}

Oscillations of the nanowire (NaugaNeedles) were measured by placing its tip in the focus of a laser beam (here, a part of the 866~nm ion cooling beam was used) and imaging its shadow onto a dual-element Si PIN photodiode \cite{gloppe14a}. The amplitude of the vibrations was strongest at the oscillator's resonance frequencies. The driven motion of the oscillator lead to modulations of the difference signal from the two elements of the photodiode. The signal was demodulated at the drive frequency $f_{\text{drive}}$ using a lock-in amplifier (Zurich Instruments HF2LI) yielding the mechanical excitation spectrum displayed in Fig. \ref{fig2}a. The observed features correspond to combined excitations of the nanowire on its tungsten holder. The measured eigenfrequencies were found to be consistent with simulations performed with the structural mechanics module of COMSOL Multiphysics \cite{COMSOL}. In these simulations, the tungsten holder geometry was modeled as a cylindrical base (diameter $d_{\text{holder}}=250$~\textmu m and length $l_{\text{holder}}=4.15$~mm) with a conical tip of length $l_{\text{tip}}=715$~\textmu m. The Ag$_{2}$Ga nanowire was modeled as a cylinder of diameter $d_{\text{nw}}=150$~nm and length $l_{\text{nw}}=16$~\textmu m placed on the tip of the holder. The $n=4$ mode used in the present experiments corresponds to a vibration with a predominant oscillation amplitude of the nanowire and minor contributions of the support. 

\subsection*{Photon correlation method}

A photon correlation method \cite{berkeland98a, bwang10a, keller15a} was employed for the detection and quantification of ion motion. This method was originally introduced for the detection of excess ion micromotion in radiofrequency traps \cite{berkeland98a} and is used here for the measurement of \textit{secular} ion motion driven by the charged mechanical oscillator. Briefly, laser-cooled ions undergoing oscillatory motion in a trap exhibit periodic variations of their fluorescence because of a continuously varying detuning of the laser frequency with respect to the cooling resonance in the frame of the ion due to the Doppler effect. 
The photon scattering rate $R(t)$ is thus modulated according to 
\begin{equation}
\label{eq:PCMfit_methods}
    R(t) = R_{0}\frac{(\Gamma_{12}/2)^2}{(\Gamma_{12}/2)^2+(\delta_0-kv_\text{max}\cos{(\omega t-\varphi)})^2}
\end{equation}\\
where $R_0$ is the scattering rate on resonance, $\Gamma_{12}$ the natural linewidth of the transition, $\delta_0$ the detuning of the cooling-laser frequency from resonance for an ion at rest, $k$ is the projection of the laser's wave vector on the direction of motion of the ion, $v_\text{max}$ is the maximum velocity of the ion during vibration, $\omega$ is the oscillation frequency, $t$ is the time and $\varphi$ is a phase determined by the initial conditions of the experiment.
The arrival times of the fluorescence photons on the PMT were measured with a time-to-amplitude converter (TAC, Stanford Research Systems SR620) in correlation with the drive of the nanowire yielding the time-resolved fluorescence curves displayed in Fig.~\ref{fig2}b. A fit of the data to Eq.~\ref{eq:PCMfit_methods} yields the velocity amplitudes $v_\text{max}$ imparted to the ion by the mechanical drive of the nanowire. Uncertainties of $v_\text{max}$ quoted include both fit and statistical errors of the measurement.

\subsection*{Theoretical modelling}

The ion-nanowire hybrid system was modelled using classical-dynamics simulations following the treatment of Ref.~\cite{fountas19a}. The nanowire was modeled as a single point charge at position $\Vec{x}_{\text{nw}}=(x_{\text{nw}},y_{\text{nw}},z_{\text{nw}})$, corresponding to the leading term of the multipole expansion of the interaction of its asymmetric charge distribution with the ion. The total electric potential $\Phi_{\text{tot}}$ experienced by an ion at position $\Vec{x}=(x,y,z)$ was thus described by the sum of the trapping potential $\Phi_{\text{trap}}$ and a Coulomb interaction:
\\
\begin{equation}
\label{eq:IApotential}
    \Phi_{\text{tot}}=\Phi_{\text{trap}}+k_{\text{c}}\frac{q_{\text{nw}}q_{\text{ion}}}{\lvert\Vec{r}\rvert}
\end{equation}\\
\\
Here, $\lvert\Vec{r}\rvert=\sqrt{(x-x_{\text{nw}})^2+(y-y_{\text{nw}})^2+(z-z_{\text{nw}})^2}$ is the effective ion-nanowire distance, $q_{\text{nw}}$ and $q_{\text{ion}}$ are the charges of the nanowire and the ion and $k_{\text{c}}$ is the Coulomb constant. The mechanical drive of the nanowire was added by substituting $z_{\text{nw}}\rightarrow z_{\text{nw}}+A\cos{(\omega_{\text{nw}}t)}$, where $A$ is the amplitude and $\omega_{\text{nw}}$ the angular frequency of the nanowire oscillation. The nanowire was assumed to vibrate in the $z$-direction in line with the direction of the piezo drive in the experiment. $\Phi_{\text{trap}}$ was expressed by a time-independent harmonic pseudopotential \cite{major05a}. \\
Taylor expansion of the interaction potential to second order yielded an approximate simulation model with equations of motions corresponding to a periodically driven harmonic oscillator (see also \cite{hunger11a}). From analysis of the Taylor-expanded potential, $v_{\text{max}}$ behaves as a function of $A$ and $d$ as \\
\begin{equation}
    v_{\text{max}}\propto(\frac{1}{d^3}-3\frac{z_{\text{nw}}^2}{d^5})A
\end{equation}\\
where $d$ is the distance between the ion-nanowire equilibrium positions. This expression was used to fit the data in Figs.~\ref{fig4}a,b. 

The equations of motion were solved numerically using the Velocity Verlet algorithm \cite{spreiter99a}. Typical integration time steps were $\Delta t=1$~ns with $1\times10^6$ steps leading to a total simulation time of 1~ms. The nanowire oscillation frequency was matched to the axial trap frequency $\omega_{\text{nw}}=\omega_{\text{z}}$ as in the experiments. The nanowire's amplitude response $A(f_\text{drive})$ to the drive frequency $f_\text{drive}$ was modeled by a Lorentzian function \cite{biedermann10a}:\\
\begin{equation}
    A(f_\text{drive})=A_0\frac{f_{\text{R}}^2/Q}{\sqrt{(f_{\text{R}}^2-f_\text{drive}^2)^2+f_\text{R}^2f_\text{drive}^2/Q^2}}
\end{equation}\\
where $f_\text{R}=422$~kHz is the $n=4$ resonance frequency and $A_0$ is the resonant amplitude response at $f_{\text{drive}}=f_{\text{R}}$. The quality factor $Q=f_{\text{R}}/\Delta f=117$ was estimated from the full-width-at-half-maximum (FWHM) $\Delta f=3.6$~kHz of the 422 kHz resonance peak (see fig. \ref{fig3}). 

\bibliography{main}% Produces the bibliography via BibTeX.

\end{document}